# A Review of Privacy Essentials for Confidential Mobile Data Transactions

Kato Mivule[1] and Claude Turner[2]

[1,2] Computer Science Department, Bowie State University, USA

[1] mivulek@students.bowiestate.edu; [2] cturner@bowiestate.edu

*Abstract—* The increasingly rapid use of mobile devices for data transaction around the world has consequently led to a new problem, and that is, how to engage in mobile data transactions while maintaining an acceptable level of data privacy and security. While most mobile devices engage in data transactions through a data cloud or a set of data servers, it is still possible to apply data confidentiality across data servers, and, as such, preserving privacy in any mobile data transaction. Yet still, it is essential that a review of data privacy, data utility, the techniques, and methodologies employed in the data privacy process, is done, as the underlying data privacy principles remain the same. In this paper, as a contribution, we present a review of data privacy essentials that are fundamental in delivering any appropriate analysis and specific methodology implementation for various data privacy needs in mobile data transactions and computation.

*Keywords—* Data privacy; data utility; anonymity; disclosure control; confidentiality; data transactions

## I. INTRODUCTION

The increasingly rapid use of mobile devices for data transaction around the world has consequently led to a new problem, and that is, how to engage in mobile data transactions while maintaining an acceptable level of data privacy and security. Therefore, in this paper, we present a review of data privacy essentials that are fundamental in delivering any appropriate analysis and specific methodology implementation for various data privacy needs in mobile data transactions and computation. However, it is necessary to distinguish between data privacy and data security, with the former dealing with confidentiality control, and the latter involves handling accessibility control. Data privacy is the procedure of protecting an individual or entity against illegal data disclosure while data security is the control of data from illegal access [1], [2]. While a dataset might be secure, it might not necessarily be private.

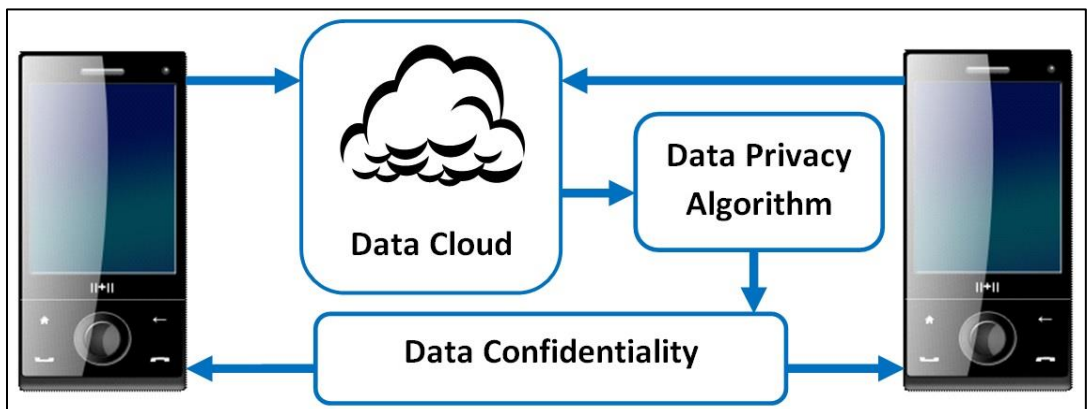

Fig 1. A privacy preserving mobile data transaction model

To exemplify this fundamental point, a house might be secured with locks to ensure access control; however, bystanders could still look inside the house from a distance if there are no curtains in the windows, thus no privacy even while access is denied to the bystanders. It is therefore crucial that a review of data privacy, data utility, the techniques, and methodologies employed in the data privacy process, is done, for suitable application in the mobile data transaction domain. In this paper, we assume that mobile devices access data through the data cloud, and as shown in Fig 1., data privacy is implemented (in the data cloud) before a response to a data request. The remainder of this paper is arranged as follows: in Section II, a presentation of data privacy and utility essentials is done. In Section III, a discussion of statistical databases and disclosure limitation control methods is done. Finally, a conclusion is given in Section IV.

## II. DATA PRIVACY AND UTILITY

*Ambiguity in defining privacy:* No precise standard definition for privacy appears in literature [3], [49]; privacy is a human and socially driven characteristic made up of human personalities such as perceptions and opinions, making the description of what data privacy is subjective and dependent on the individual or entity and what personal data they are willing to disclose [48]. Therefore, it is not practical to craft a universal data privacy solution, however, various individuals and entities will have differing data privacy requirements and thus customized data privacy solutions [4], [49]. *Personally Identifiable Information (PII):* is any data about an individual or entity that could be used to generate the complete identity of that individual or entity, for example social security and passport numbers [5], [6]. *The PII description problem:* A precise and complete definition of PII is still problematic as researchers have shown that non-PII attributes can still be used in conjunction with other data to link and reconstruct the identities of individuals [7]. Additionally, legal scholars have observed that there exists no consistent explanation of PII in information privacy law and that current descriptions are inadequate due to the ever changing landscape of what constitutes PII [8]. Furthermore, what constitutes PII in one geographic region might be different from another, only adding to the intricacy of the problem; for example a zip code could be viewed as linkable information in the USA but as of this date, zip codes are none existent in the nation of Uganda in Africa [9]. Therefore the need to protect what makes up PII largely depends on what the individual considers as sensitive information [10] and, as such, data utility will have to account for such descriptions. Any upcoming architectures of privacy will necessitate data gleaners to act in ways that are in line with consumer views of what privacy is thus giving individuals leverage to dictate terms of privacy and hence data utility [11]. *Data Privacy Verses Data Utility:* Data utility verses privacy is the notion of how beneficial a privatized dataset is to a user of that published privatized dataset [43], [44]. During the data privacy process, PII data is removed and noise is added to ensure privacy. However, the utility (usefulness) of data diminishes during the data privacy process, rendering the privatized dataset meaningless to the user of that dataset as more PII is removed and sensitive data altered to further provide concealment. Equilibrium between data privacy and utility requirements is persistently pursued [45]; however, researchers have found that attaining such a goal of optimal data privacy while not diminishing data utility is a continual intractable challenge [46].

### III. STATISTICAL DATABASES AND DISCLOSURE CONTROL METHODS

Statistical disclosure control (SDC) also known as data de-identification, data anonymization, and data sanitization, is a procedure that removes PII data or transforms sensitive data to a point that when that data is published, an individual's identity or an entity's sensitive information cannot be exposed [5], [12], [13], [14]. Statistical disclosure control techniques are categorized in two groups, as perturbative and non-perturbative. Perturbative SDC methods transform the original data by generally using noise addition methods to conceal sensitive data while non-perturbative techniques do not transform the original data but rather suppress or generalize sensitive data to provide concealment [1]. Knowing the difference between the various categories of SDC methods is essential, as different datasets would require specialized data privacy techniques for confidentiality. For instance, continuous datasets might work well with perturbative methods while non-perturbative methods would best be prescribed for categorical datasets.

*A. Statistical databases*

Statistical databases are datasets made available to the public and do not change. If the publisher of that dataset has newer entries, then another static dataset is published with new entries in the updated version [15]. Statistical databases can be categorized into two groups as microdata and macrodata: *Microdata* refer to non-summarized statistical data about an individual or entity, with each row in the microdata representing the full record of the individual or entity while *Macrodata* refers to summarized or aggregated data about a group of individuals or entities, with each row representing the aggregate data of a group of individuals or entities [16], [17]. During the data privacy process, a look at what statistical databases are composed of would be useful in specifying which attributes to conceal or reveal [1], [18]: *Attributes:* attributes are columns, column titles, or column names. *PII attributes:* these are attributes with information that can distinctively reveal the identity of an individual. *Quasi-attributes:* these are columns that don't store PII information but can be used in conjunction with other external information to reveal the identity of an individual [9]. *Confidential or sensitive attributes:* when looking at microdata**,** these are columns or attributes that do not contain PII data and are not quasi-attributes but store data that is sensitive; examples of such sensitive data could include HIV and cancer diagnosis [9]. *Non-confidential attributes:* these are columns or attributes that do not reveal any sensitive information. *Categorical and continuous data types:* It is essential to know the type of statistical data to handle during the data privatization process, as each type of statistical data would require a different data privatization algorithm and thus different measure of data utility [9]. *Categorical data:* is representative of values for which we cannot perform mathematical operations but can be categorized into groups, for example male, female, car models, categories of birds, use frequencies to count data, and Non parametric statistical techniques for statistical analysis; while, *Continuous data* is representative of values for which we can perform mathematical calculations such as age, hours worked, salary, and miles travelled. Parametric statistical techniques used for statistical analysis on continuous data [9], [18], [19], [20].*Univariate and multivariate datasets: Univariate datasets* are made up of one variable and thus statistical observation is made on a single variable while *Multivariate datasets,* on the other hand, are composed of two or more variables, as such, statistical observation is made on two or more variables [21], [22].

| PII attributes | | | Quasi attributes | | | Non-sensitive attributes | | Sensitive attributes | |
|---|---|---|---|---|---|---|---|---|---|
| FName | LName | SSN | Date of Birth | Age | Gender | Zip Code | City | Income | Diagnosis |
| Tom | Blue | 000-000-0001 | 02-20-1960 | 53 | M | 55555 | Music City | 20000.00 | Cancer |
| Nancy | Indigo | 000-000-0002 | 01-01-1968 | 48 | F | 12345 | Blue Ridge | 30000.00 | Cancer |
| Jane | White | 000-000-0003 | 02-22-1970 | 43 | F | 12345 | Chicago | 50000.00 | HIV |
| Green | Black | 000-000-0004 | 12-12-1973 | 40 | M | 55555 | Denver | 60000.00 | HIV |
| Fred | Blue | 000-000-0005 | 12-11-1972 | 41 | M | 21111 | Movie City | 75000.00 | Diabetic |
| Anne | White | 000-000-0006 | 01-01-1969 | 47 | F | 31111 | Fame City | 20000.00 | Cancer |
| Gregory | Brown | 000-000-0007 | 04-20-1961 | 52 | M | 41111 | Music City | 45000.00 | HIV |
| Keith | Ingido | 000-000-0008 | 029-20-1962 | 51 | M | 12345 | Blue Ridge | 80000.00 | Diabetic |
| Freda | Yellow | 000-000-0009 | 08-22-1970 | 43 | F | 31111 | Fame City | 15000.00 | Cancer |
| Kim | Black | 000-000-0010 | 11-12-1970 | 43 | F | 31000 | Camp City | 20000.00 | HIV |

Fig 2. An illustration of PII, quasi, non-sensitive, and sensitive attributes.

*Parametric and nonparametric techniques: Parametric methods* refer to statistical measurements based on the normal distribution of a sample data, and characterized by the mean, variance, and independent observations. Parametric methods are employed for numerical data. *Non-parametric statistical methods,* on the other hand, do not depend on normal distribution, variance, and independence of sample data but are based on few suppositions; non-parametric methods would be considered for analysis of categorical data [23], [24].

| Microdata | | | | | | Macrodata | |
|---|---|---|---|---|---|---|---|
| Multivariate - variables X and Y | | | | Univariate - Single variable Z | | | |
| SCHOOL X | | SCHOOL Y | | | | | |
| Student | GPA | Student | GPA | GPA | | Age Group | Grouped Income |
| Blue | 3.4 | Jame | 2.1 | 2.1 | | 16 -19 | 5000 - 15000 |
| Indigo | 3.3 | Petre | 2.4 | 2.4 | | 20 -30 | 10000 - 30000 |
| White | 3.7 | Jane | 3.7 | 3.7 | | 31 - 40 | 40000 - 50000 |
| Black | 2.9 | Jons | 2.9 | 2.9 | | 41 - 50 | 50000 - 60000 |
| Blue | 3.1 | Jude | 3.4 | 3.4 | | 51 - 60 | 60000 - 120000 |
| White | 2.1 | Sally | 4.0 | 4.0 | | 61 - 70 | 60000 - 120000 |
| Brown | 2.3 | Ken | 2.3 | 2.3 | | 71 - 80 | 50000 - 80000 |
| Ingido | 4.0 | Jim | 4.0 | 4.0 | | 81 - 90 | 20000 - 30000 |
| Yellow | 3.4 | Jon | 3.4 | 3.4 | | 91 - 100 | 20001 - 30000 |
| Black | 3.2 | Bob | 3.2 | 3.2 | | 101 - 120 | 20002 - 30000 |
| Categorical Data | | | | Continuous Data | | | |

Fig 3. An illustration of multivariate, univariate, categorical, continuous, microdata, and macrodata

*Horizontal and vertical data partitions:* In data mining, large datasets are always split into reduced portions of data for manageable and efficient processing. *Horizontal partitioning* involves the splitting of a large dataset by rows into isolated sets of smaller rows, with each row of data containing all attributes for each record. *Vertical partitioning* on the other hand involves the splitting of the large dataset by columns, along attributes into separate sets of smaller columns; as such an attribute of data retains all the values of each row for that particular attribute. However, the data for every attribute in the record or row is not retained [25], [26].

| Horizontal partition along rows | | | | | | Vertical partition of data along columns | | | |
|---|---|---|---|---|---|---|---|---|---|
| FName | LName | SSN | Date of Birth | Age | Gender | LName | Diagnosis | Age | Income |
| Tom | Blue | 000-000-0001 | 02-20-1960 | 53 | M | Blue | Cancer | 53 | 20000.00 |
| Nancy | Indigo | 000-000-0002 | 01-01-1968 | 48 | F | Indigo | Cancer | 48 | 30000.00 |
| Jane | White | 000-000-0003 | 02-22-1970 | 43 | F | White | HIV | 43 | 50000.00 |
| Green | Black | 000-000-0004 | 12-12-1973 | 40 | M | Black | HIV | 40 | 60000.00 |
| | | | | | | Blue | Diabetic | 41 | 75000.00 |
| | | | | | | White | Cancer | 47 | 20000.00 |
| FName | LName | SSN | Date of Birth | Age | Gender | Brown | HIV | 52 | 45000.00 |
| Fred | Blue | 000-000-0005 | 12-11-1972 | 41 | M | Ingido | Diabetic | 51 | 80000.00 |
| Anne | White | 000-000-0006 | 01-01-1969 | 47 | F | Yellow | Cancer | 43 | 15000.00 |
| Gregory | Brown | 000-000-0007 | 04-20-1961 | 52 | M | Black | HIV | 43 | 20000.00 |
| Keith | Ingido | 000-000-0008 | 029-20-1962 | 51 | M | Black | HIV | 45 | 21000.00 |
| Freda | Yellow | 000-000-0009 | 08-22-1970 | 43 | F | White | Cancer | 32 | 34000.00 |
| Kim | Black | 000-000-0010 | 11-12-1970 | 43 | F | Blue | Cancer | 55 | 35000.00 |

Fig 4. Illustration of a horizontal and vertical partition of micro data.

B. *Data privacy techniques*

Data privacy methods in SDC can be characterized into two groups: *Non-perturbative* approaches that do not modify the values of the original data during the privacy process. *Perturbative methods,* in which values of the original data are transformed, masked, or camouflaged for confidentiality [1]. While a wide range of data privacy methods have been proposed, in this study, we examine non-perturbative techniques such as *k*-anonymity, *l*-diversity, suppression, and generalization. *Suppression:* Suppression is a data privacy enforcing mechanism in which PII and sensitive data or values are deleted or removed at the cell level from the dataset. An example includes deleting the highest and lowest income values from a dataset that might stand out as unique. Often suppression is used in conjunction with methods such as generalization and *k*-anonymity [9, 13].

Fig 5. An illustration of generalization, suppression, k-anonymity, and l-diversity

*Generalization:* Generalization is a data privacy technique in which a group of unique values in that same attribute is given a single value. Generalization of values follows a domain generalization hierarchy (DGH), which is the level at which to generalize a value. For example, we might begin with a birthdate attribute in which $B_1 = \{1961\text{-}01\text{-}01\} \rightarrow B_2 = \{1961\text{-}01\} \rightarrow B_3 = \{1961\}$, thus blanketing all values in the birthdate attribute with one single value, *1961*, and where $B_1,...,B_n$ are the DGH levels of birthdate *B* [9]. *K-anonymity*: *k*-anonymity enforces data privacy by requiring that all values in the quasi-attributes be repeated *k* times, such that $k > 1$, so as to provide confidentiality, making it harder to uniquely distinguish individuals. However, *k*-anonymity incorporates generalization and suppression methods on unique values so as to achieve $k > 1$ [9], [27] [28]. For example, if we had a zip code attribute, $z = \{20001, 20002, 20001, 20005, 20005\}$, k-anonymity would require that any given zip code in set *z* be repeated $k > 1$ times in the privatized dataset $z' = \{20001, 20001, 20005, 20005\}$; using suppression, we deleted the unique zip code *20002* and thus we have *20001* and *20005* repeating twice such that $k = 2$, and satisfying the *k*-anonymity requirement of $k > 1$. *l-Diversity:* While *k*-anonymity requires that values in quasi-attributes be repeated at least $k > 1$ times to provide confidentiality, researchers have shown that even while such quasi-attribute and sensitive attribute values might be repeated $k > 1$ times, this is a major weakness as an attacker only needs to look at the generalized sensitive attributes to reveal sensitive information about any individual [29]. To overcome this problem, Machanavajjhala, et al., (2007) proposed *l*-diversity as an addition to *k*-anonymity by requiring that for any data privacy procedure to satisfy *l*-diversity, it has to first meet the requirements for *k*-anonymity for all quasi-attributes, and then ensure that there are *l* diverse values in the sensitive attributes. Since *l*-diversity works in conjunction with *k*-anonymity, researchers have found that achieving *l*-diversity is also NP-hard and thus intractable [30], [31].

C. *Perturbation Techniques*

*Noise addition*: a generation of random values selected from a normal distribution with zero mean and a very small standard deviation is done, and then added to sensitive numerical attribute values to ensure privacy. The general expression of noise addition is [32],[ 49]:

$$X + \varepsilon = Z \qquad (1)$$

*X* is the original continuous dataset and *ε* is the set of random values (noise) with a distribution $e \sim N(0, \sigma^2)$ that is added to *X*, and finally *Z* is the privatized dataset [32], [49]. *Multiplicative noise*: a generation of random values with mean $\mu = 1$ and variance $\sigma^2$, is done and then multiplied to the original values, the result is then published as the privatized dataset. A formal description of multiplicative noise is as follows:

$$X_j \varepsilon_j = Y_j \qquad (2)$$

$X_j$ is the original value; $\varepsilon_j$ is the random values with mean $\mu = 1$ and variance $\sigma^2$; and finally, $Y_j$ is the privatized dataset after multiplying the contents of $X_j$ and $\varepsilon_j$ [32], [49]. *Logarithmic multiplicative noise*: an adaptation of the multiplicative noise technique that makes a logarithmic adjustment of the original values as shown below:

$$lnX_j = Y_j \tag{3}$$

Random values $\varepsilon_j$ are then created and added the logarithmic adjusted values, $Y_j$, finally producing the privatized values $Z_j$ as shown below:

$$Y_j + \varepsilon_j = Z_j \tag{4}$$

The original values are represented by $X_j$; $Y_j$ symbolizes the logarithmic perturbed values; $Z_j$ is the privatized data values [32], [49].

*Differential Privacy (DP)*: data privacy is enforced by adding Laplace noise to query responses from the database in such a way that the users of the database cannot differentiate if a specific value has been changed in that database [32], [33]. Two databases $D_1$ and $D_2$ are viewed as undistinguishable if they are different by only one element such that $\boldsymbol{D_1 \Delta D_2 = 1}$. Any data privacy technique $\boldsymbol{q_n}$ that grants privacy, meets the requirements of $\boldsymbol{\varepsilon}$-*differential privacy* if the probability of the outcome (query responses) to that same query executed on database $D_1$ and then executed on database $D_2$ should be alike, and satisfying the following condition [32], [33]:

$$\frac{P[q_n(D_1) \in R]}{P[q_n(D_2) \in R]} \leq e^{\varepsilon} \tag{5}$$

Where $D_1$ and $D_2$ are the two databases, $P$ is the probability of the Laplace noise induced query responses $D_1$ and $D_2$ correspondingly. $q_n()$ is the privacy technique (perturbation); $q_n(D_1)$ is the privacy technique on query responses from database $D_1$; $q_n(D_2)$ is the privacy technique on query results from database $D_2$; $R$ is the Laplace noise induced query responses from the databases $D_1$ and $D_2$ respectively; and $e^{\varepsilon}$ is the small exponential $\varepsilon$ epsilon value. To implement DP, the following steps are done: The max difference is calculated, $\Delta f$ is the max difference (most influential observation) [32], [33], [34]:

$$\Delta f = Max|f(D_1) - f(D_2)| \tag{6}$$

Laplace noise between (0, b) is generated and added to $f(x)$, the original query response, such that:

$$b = \frac{\Delta f}{\varepsilon} \tag{7}$$

$$Differential\ private\ data = f(x) + Laplace(0, b) \tag{8}$$

*Data Swapping:* Data swapping is a data privacy technique that involves substituting values of sensitive variables with other variables in the same dataset while maintaining the original frequencies and statistical properties [35], [36]. Data swapping has been generally used by the US Census Bureau [37]; yet researchers have noted that data swapping distorts data because of the alterations of the joint distributions between swapped and non-swapped attributes [38]. For example, if given two employees with age and income values such that, *A* = *{20, 10000}*, and *B* = *{30, 30000}* respectively; during data swapping, data privacy specialists would exchange the income and age of employee *A* with *B* and likewise the age and income values of *B* are assigned to *A*, such that the privatized dataset becomes *A* = *{30, 30000}* and *B* = *{20, 10000}*. *Synthetic data sets:* In synthetic data generation, an original set of tuples is replaced with a new set of look-alike tuples but while still preserving the statistical properties of the original data values [1]. Synthetic data generation falls in two major categories, fully synthetic and partially synthetic. Fully synthetic datasets are unreal or pseudo datasets created by replacing values in the original dataset with imputed unknown data values that retain the same statistical characteristics as in the original dataset but totally hide any sensitive or private information [39]. Other mathematically based SDC methods include the following [13],[40], [41], [42]: *Top-coding and bottom-coding:* top-coding and bottom-coding involves publishing data values based on the high-end or low-end of a given value. *Recording:* in the recording data privacy approach, individual data values are allocated to group values or ranges of values instead of publishing the exact values in the dataset. *Rounding:* in the rounding technique data values in the original dataset are replaced with rounded values up or down on a set of attributes. *Blank and impute:* in the blank and impute technique, sensitive data values are deleted and then replaced with values that have been statistically modelled or with values that are similar to other values in the same dataset. *Blurring:* involves determining a given value randomly or selectively and then replacing that value with an average value. *Cryptographic Techniques:* involve the use of secure computation or encryption to release data or query responses over multiple databases without revealing any data other than the answer to a particular query, ensuring that privacy-threatening data mining is prevented.

IV. EXPERIMENT

As an illustration, we set up an experiment to implement noise addition on a data set for privacy. We used the publically available Iris Fisher data set from the UCI repository containing 150 data items [47]. We generated random noise with a distribution between the mean and standard deviation of each attribute, that is the sepal length, sepal width, petal length, and petal width. As can be shown in Fig 5., the distribution of the petal length in the original iris data set is clearly separable, however, after noise is added to the data set, as shown in Fig 6., the petal length becomes difficult to separate, thus more distortion. While this is a limited experiment for demonstration purposes, it clearly shows that while privacy might be achieved on a data set by adding more noise, data utility (the usefulness of a data set) is diminished. Therefore the underlying problem of privacy versus utility remains a challenge – the more private a data set is, the less data utility will be achieved on that data set.

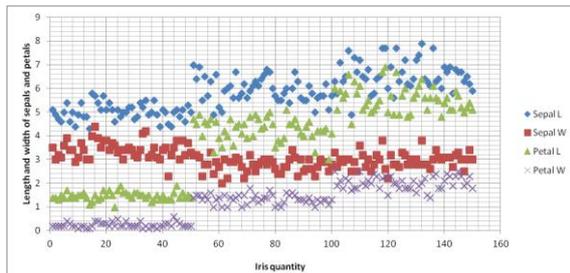

Fig 6. Original Iris data scatter plot distribution.

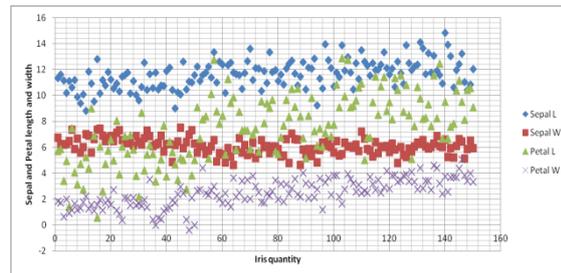

Fig 7. Privatized Iris data scatter plot distribution.

V. CONCLUSIONS

We have presented a review of data privacy essentials that are fundamental in delivering any appropriate analysis and specific methodology for various data privacy needs in mobile data transactions and computation. While a number of data privacy algorithms have been proposed, the problem of data privacy versus data utility is still a challenge. Finding a balance between data privacy and utility needs requires trade-offs. Yet still, implementation of such data privacy enhancing algorithms in the mobile computing domain remains a challenge that researchers still have to tackle, even as mobile computing becomes the standard means of transacting in data around the globe.

ACKNOWLEDGMENT

Special thanks to Dr. Claude Turner, the entire computer science department at Bowie State University, and the HBGI grant from the department of education that made this work possible.